\documentclass[aps,prl,twocolumn,showpacs,floatfix,superscriptaddress]{revtex4-2}%
\usepackage[pdftex                     
           ,pagebackref=false          
           ,colorlinks=true           
           ]{hyperref}
\usepackage{amscd}
\usepackage{amssymb}
\usepackage{amsthm}
\usepackage{amsfonts}
\usepackage{amstext}
\usepackage{textcomp}
\usepackage[final]{graphicx}
\usepackage{dcolumn}
\usepackage{bm}
\RequirePackage{lineno}
\usepackage[usenames]{color}
\usepackage[normalem]{ulem}
\usepackage{epsfig} 
\usepackage{epstopdf}
\usepackage{url}
\usepackage{mathtools}
\usepackage{times}

\hypersetup{linkcolor= blue,           
            citecolor= blue,            
            urlcolor=blue}            

\begin{document}

\title {Possible Eliashberg-type superconductivity enhancement effects in a two-band superconductor MgB$_{2}$ driven by narrow-band THz pulses}

\author{Sergei\,Sobolev}
\author{Amon\,Lanz}
\affiliation{Institute of Physics, Johannes Gutenberg University, 55128 Mainz, Germany}
\author{Tao\,Dong}
\affiliation{Institute of Physics, Johannes Gutenberg University, 55128 Mainz, Germany}
\affiliation{International Center For Quantum Materials (ICQM), Peking University, Beijing 100871, China}
\author{Amrit\,Pokharel}
\affiliation{Institute of Physics, Johannes Gutenberg University, 55128 Mainz, Germany}
\author{Viktor\,Kabanov}
\affiliation{Josef Stefan Institute, 1000 Ljubljana, Slovenia}
\author{Tie-Quan Xu}
\author{Yue Wang}
\author{Zi-Zhao Gan}
\affiliation{Applied Superconductivity Center and State Key Laboratory for Mesoscopic Physics, School of Physics, Peking University, Beijing 100871, China}
\author{L.Y.\,Shi}
\author{Nan-Lin\,Wang}
\affiliation{International Center For Quantum Materials (ICQM), Peking University, Beijing 100871, China}
\author{Alexej\,Pashkin}
\author{Ece\, Uykur}
\author{Stephan\,Winnerl}
\author{Manfred\, Helm}
\affiliation{Helmholtz-Zentrum Dresden-Rossendorf, 01314 Dresden, Germany}
\author{Jure\,Demsar}
\affiliation{Institute of Physics, Johannes Gutenberg University, 55128 Mainz, Germany}
\email{demsar@uni-mainz.de}


\begin{abstract}
We study THz-driven condensate dynamics in epitaxial thin films of MgB$_{2}$, a prototype two-band superconductor (SC) with weak interband coupling. The temperature and excitation density dependent dynamics follow the behavior predicted by the phenomenological bottleneck model for the single-gap SC, implying adiabatic coupling between the two condensates on the ps timescale. The amplitude of the THz-driven suppression of condensate density reveals an unexpected decrease in pair-breaking efficiency with increasing temperature - unlike in the case of optical excitation. The reduced pair-breaking efficiency of narrow-band THz pulses, displaying minimum near $\approx0.7$ T$_{c}$, is attributed to THz-driven, long-lived, non-thermal quasiparticle distribution, resulting in Eliashberg-type enhancement of superconductivity, competing with pair-breaking.

\end{abstract}
\maketitle

Light-induced manipulation of symmetry-broken ground states like
superconductors, density waves or magnetic materials has been at the forefront
of modern solid-state physics research for over a decade. While excitation
with femtosecond optical pulses at photon energies much larger than energies
of gaps or relevant bosonic excitations normally results in a light-induced
suppression of the order parameter, excitation of selected lattice vibrations
or resonant electronic transitions may result in metastable states that cannot
be reached by quasi-thermal pathways. Examples of such include modulation of
optical gap in CH$_{3}$NH$_{3}$PbI$_{3}$ by THz driving of optical phonons~\cite{Heejae}, possible light-induced superconductivity at temperatures above the superconducting critical temperature~\cite{Mitrano,Shimano,Cavalleri}, switching to a hidden state in 1T-TaS$_{2}$~\cite{Dragan}, generation of spin-density-wave order at high temperatures in BaFe$_{2}$As$_{2}%
$~\cite{Kim}, or inducing ferroelectricity in SrTiO$_{3}$~\cite{Nova}.

While most of these approaches are based on non-adiabatic effects, light-induced enhancement of superconductivity (SC) in conventional BCS superconductors has first been
demonstrated in 1960's \cite{Clarke1,Clarke2,Tinkham,DemsarReview}.
The fascinating enhancement of critical temperature, critical current and the
superconducting gap was observed in clean type-I superconductors under
continuous illumination with electromagnetic radiation at sub-gap frequencies \cite{Clarke1,Clarke2}. While sub-gap excitation does not lead to condensate depletion and thus conserves the number of quasiparticles (QPs), it may result in a non-thermal QP distribution via excitation of thermally excited QPs away from the gap edge. As the low-lying QPs are much more effective in inhibiting pairing correlations than high-energy QPs, the non-thermal QP distribution results in the enhancement of SC, as demonstrated by Eliashberg by solving the self-consistent BCS gap equation under such non-equilibrium
\cite{Eliashberg1,Eliashberg2,CurtisCavity}. Refinements of the Eliashberg model, considering that the QP recombination rate is increasing with increasing QP energy, suggested that even an overall reduction of QP density can be realized \cite{Scalapino,Schoen}.

As such effects rely on a non-thermal QP distribution, experimental studies with continuous microwave illumination were largely limited to materials with large electron mean free paths, and the resulting enhancements of the gap, critical current, and critical temperature were of the order of 1\% \cite{Tinkham,DemsarReview}. However, even in NbN, a prototype dirty-limit SC, signatures of gap enhancement were observed on a 50~ps timescale, when exciting with narrow-band THz pulses \cite{Beck1}. To investigate such transient phenomena further, MgB$_{2}$, a prototype two-band BCS superconductor\ with the highest critical temperature (T$_{c}\approx 40\ K$) among metallic phonon-mediated BCS superconductors (at ambient
pressure) presents an interesting alternative. MgB$_{2}$ is
characterized by two distinct types of electronic bands crossing the Fermi energy, $E_{f}$, the quasi-two-dimensional hole-like $\sigma-$bands and the three-dimensional $\pi-$bands \cite{Liu,Xi2008}. As demonstrated by tunnelling
\cite{Szabo2001,Iavarone2002}, photoemission \cite{Souma}, and Raman spectroscopy \cite{Blumberg2007}, the superconducting state is characterized by two distinct isotropic gaps with $\Delta_{\pi}\approx2$~meV and $\Delta_{\sigma}\approx7$~meV
\cite{Szabo2001,Iavarone2002,Souma,Blumberg2007,Prozorov_Symm} in the $\pi$- and $\sigma$-bands, respectively. The presence of a single $T_{c}$ with two well-distinguished gaps\ suggests a weak, yet not negligible, interband coupling \cite{Liu,Xi2008}. The low intraband scattering rates
\cite{THG_Tao,Fiore2022} and the relatively large value of $\Delta_{\sigma}$
make MgB$_2$ an ideal test-system to study SC enhancement effects driven by
narrow-band THz pulses.

In this Letter, we present systematic studies of condensate dynamics in clean
epitaxial MgB$_{2}$\ thin films {\cite{films,THG_Tao}}, following excitation
with narrow-band THz pulses, with frequencies between $2\Delta_{\pi}/h$ and $2\Delta_{\sigma}/h$. By studying the temperature ($T$) and absorbed energy
density ($A$)\ dependence of the THz-driven gap suppression and recovery dynamics, and comparing the results with those obtained by near-infrared (NIR) pumping at 1.55~eV, we provide evidence of long-lived Eliashberg SC enhancement effects in MgB$_{2}$. These increase with increasing $T$, reflecting an increase in the density of thermally excited QPs, until reaching the temperature where also $2\Delta_{\sigma}(T)$ drops below the THz photon energy. 

Time-resolved studies of the dynamics following excitation with THz pulses were performed at the free electron laser (FEL) facility in Helmholtz-Zentrum Dresden-Rossendorf, providing intense picosecond narrow-band (bandwidth $\Delta\nu_{\text{FEL}}\approx30$~GHz) THz pulses with
tunable frequency $\nu_{\text{FEL}}$ at a repetition rate of 13~MHz \cite{Elbe, Suppl}. We investigated {high-quality epitaxial (001) single-crystalline MgB$_{2}$\ thin films grown on MgO (111) }substrate {by hybrid physical-chemical vapor deposition \cite{films}}. Films, with thicknesses of 15 and 30~nm on 5${\times}$5~mm$^{2}$ substrates were characterized by X-ray diffraction and charge transport measurements~\cite{THG_Tao}. Experiments were carried out in a single-color pump-probe configuration using $\nu_{\text{FEL}}=1.5$, $2.1$~and $2.7$~THz, where $2\Delta_{\pi}<h\nu_{\text{FEL}}<2\Delta_{\sigma}$. For reference, we performed complementary studies of dynamics driven by 50 fs NIR pulses. The high repetition rate of the FEL results in continuous heating of the sample, which is taken into account~(all the quoted temperatures below take continuous heating into account \cite{Suppl}). As critical temperatures of samples investigated in different beamtimes varied between $\approx32$~K (15~nm film) and\ $\approx36$~K (30~nm film) we present the $T$-dependent data as a function of reduced temperature $T/T_{c}$. We focus on data with  $\nu_{\text{FEL}}=2.7$~THz, providing the highest dynamic range and signal-to-noise ratio.

\begin{figure}[ptb]
\includegraphics[width=1\columnwidth]{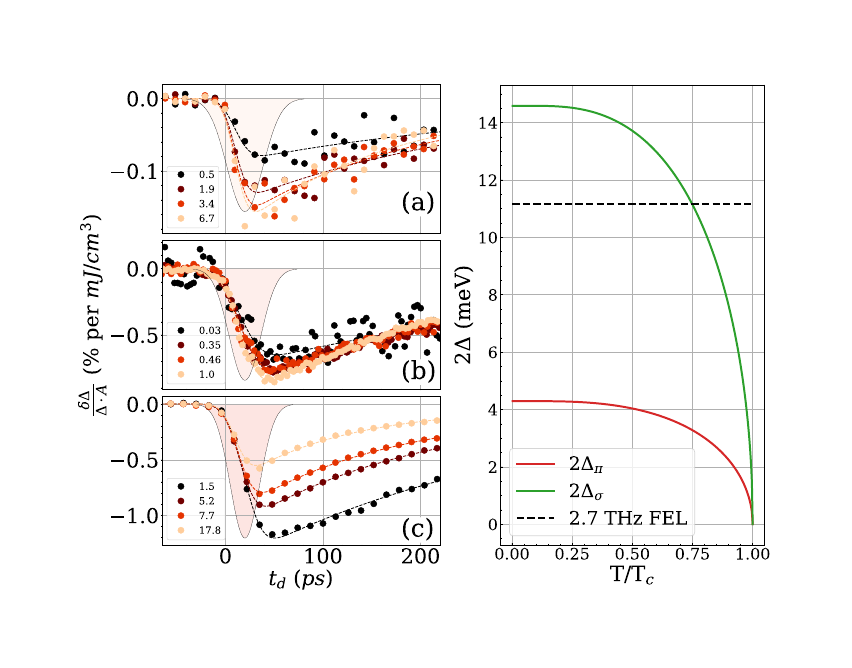}
\caption{Evolution of the THz-driven suppression of the effective gap, normalized to the absorbed energy density, $\frac{\delta
\Delta(t_{d})}{\Delta(T)\cdot A}$ recorded at the cryostat temperature T = 4.2~K with (a) ~1.5~THz, (b)~2.1~THz and (c)~2.7~THz pulses. The experimental datapoints are represented by solid symbols (absorbed energy densities, $A$ are in mJ/cm$^3$) while dashed lines represent the fits to the data using Eq.(1). The shaded regions represent the cross-correlations of THz pulses. The normalized gap suppression amplitudes are of
the same order of magnitude for all three $\nu_{\text{FEL}}$. While the relative errors in calculated $A$ for each $\nu_{\text{FEL}}$ are negligible, the absolute values of $A$\ are correct within a factor of 2 due to the large reflectivity of MgB$_{2}$ and the FEL mode quality. (d) Approximate $T-$dependence of gaps $2\Delta_{\pi}$ and $2\Delta_{\sigma}$ compared to h$\nu_{\text{FEL}}=2.7$~THz (dashed horizontal line).}%
\label{traces_freqs}%
\end{figure}

Figure \ref{traces_freqs} presents selected traces of the induced suppression and recovery of superconductivity, recorded at different $\nu_{\text{FEL}}$.
As elaborated in Supplemental Material \cite{Suppl}, these traces present the evolution of the \textit{effective} superconducting gap, $\Delta$, as a function of time delay $t_{\mathrm{d}}$. More precisely, we plot $\frac
{\delta\Delta(t_{\mathrm{d}})}{\Delta(T)}$, where $\delta\Delta(t_{\mathrm{d}%
})\equiv\Delta(t_{\mathrm{d}})-\Delta(T)$, and $\Delta(T)$ is the equilibrium value of the \textit{effective} gap at the sample temperature $T$. $\delta\Delta(t_{\mathrm{d}})$ is extracted from the recorded dynamics of the film transmission, $\delta Tr(t_{\mathrm{d}})$, and equilibrium
$Tr(T)$, assuming the so-called $T^{\ast}-$model \cite{Parker,comment}. This model, applied to describe the time evolution of the SC state in NbN \cite{Beck1}, assumes that in nonequilibrium QPs, the condensate(s) and the high frequency ($\hbar\omega>2\Delta$) bosonic excitations (in this case high-frequency acoustic phonons, HFP) are characterized by an elevated effective temperature $T^{\ast}$. The low energy ($\hbar\omega<2\Delta$) acoustic phonons thereby remain at the base temperature.

The effective gap description, where $\Delta(t)\propto$ $\Delta
_{\sigma}(t)\propto\Delta_{\pi}(t)$, is supported by the long timescales of
the dynamics and the fact that the absorbed energy density required to
suppress SC, $A_{\sup}$, matches the total condensation energy,
$E_{c}$, of MgB$_{2}$ \cite{Suppl}. Since
$E_{c}=\frac{1}{2}N_{\pi}(0)\Delta_{\pi}^{2}+\frac{1}{2}N_{\sigma}%
(0)\Delta_{\sigma}^{2}$ and the normal state densities of states in the $\pi-$
and $\sigma-$band, $N_{\pi,\sigma}(0)$, are comparable \cite{Liu,Xi2008},
$\approx90\%$ of $E_{c}$ stems from pairing in the $\sigma-$band. Thus,
$A_{\sup}\approx E_{c}$ implies that the two gaps, $\Delta_{\pi}$,
$\Delta_{\sigma}$, follow the same dynamics (beyond the resolution limit of
$\approx20$~ps) despite the fact that for excitation with $h\nu_{\text{FEL}}<2\Delta_{\sigma}$
only pairs in the $\pi-$band can directly be broken. The underlying fast thermalization between the two condensates likely proceeds through their coupling via the Leggett mode \cite{Blumberg2007,Giorgianni2019leggett}.

Since excitation with h$\nu_{\text{FEL}}>$ 2$\Delta_{\pi}$ does results in Cooper pair breaking, with $\delta\Delta$ proportional to $A$, dividing the relative gap suppression by $A$, and analyzing $\frac
{\delta\Delta(t_{\mathrm{d}})}{\Delta(T)A}$ (in $\%$ per$\ \frac{mJ}{cm^{3}}%
$), allows us to emphasize the non-linear behavior of the THz-driven gap
dynamics. Panels (a)-(c) of Figure~\ref{traces_freqs} show $\frac{\delta\Delta(t_{\mathrm{d}%
})}{\Delta(T)A}$, recorded at $T=4.2~K$ for different $A$ and 
$\nu_{\text{FEL}}=1.5$~THz (a), $2.1$~THz (b) and $2.7$~THz (c). The induced reduction of $\Delta$ is followed by the recovery of SC order on the timescale between $\approx50$~ps and several 100~ps, depending on $T$ and $A$.
We fit the traces using a model considering that pair-breaking and recovery are exponential processes \cite{nature09539}

\begin{equation}
\frac{\delta\Delta(t_{\mathrm{d}})}{\Delta(T)A}=H(t_{\mathrm{d}}%
,\tau_{\mathrm{rec}},\sigma)\times(Ce^{-\frac{t_{d}}{\tau_{\mathrm{rec}}}}+B).
\end{equation}

Here $\tau_{\text{rec}}$ is the SC state recovery time, $C$~is
the amplitude of the normalized gap suppression, and $B$~accounts for the residual gap suppression at large time delays (bolometric response). The rise-time of the recorded transient, $\sigma$, is accounted for by a Heaviside function
convoluted with the Gaussian with pulse-width $\sigma$, $H(t_{\mathrm{d}},\tau_{\mathrm{rec}},\sigma)$~\cite{nature09539}. Here, $\sigma$ can be larger than the resolution limit of $\sqrt{2}%
\tau_{\text{FEL}}$, where $\tau_{\text{FEL}}$ is the duration of the THz pulse.

Several observations can be made by inspecting the traces in
Fig.\ref{traces_freqs} (a)-(c): i) $\tau_{\text{rec}}$ is similar for
different $\nu_{\text{FEL}}$ and depends on $T$ and $A$; ii) the rise time
$\sigma$ depends on $\nu_{\text{FEL}}$, being nearly resolution limited at $\nu_{\text{FEL}}=1.5$~THz while increasing for $\nu_{\text{FEL}}\geq 2.1$~THz \cite{Suppl}; iii) over a large range of excitation densities ($\approx2$ orders of magnitude) the gap suppression amplitude displays a non-linear dependence
on $A$.

$\tau_{\text{rec}}$ and $\sigma$ display similar temperature and excitation density dependence as in the case of NIR excitation \cite{Suppl,JDMgB2}. Further, their
dependences follow the same trend as in a single-gap BCS superconductor NbN \cite{Beck1}, accounted for by the phenomenological boson bottleneck model
\cite{RT,DemsarReview}. In this model the delayed suppression of the condensate reflects the pair-breaking by bosonic excitations, while its recovery is governed by the decay of HFP population \cite{RT,DemsarReview}. In experiments with narrow-band THz excitation, such a delayed pair-breaking has not been resolved before \cite{nbn_enhan,BeckPCCO}. Since this process is clearly resolved only for $\nu_{\text{FEL}}\gtrsim2.1$ THz
(see Supplemental Material \cite{Suppl}), we speculate the process originates from pair-breaking by HFP ($\hbar\omega>2\Delta_{\pi}$) that can be created at
higher $\nu_{\text{FEL}}$ since $\nu_{\text{FEL}}\gtrsim2.1$ THz
$\gtrsim4\Delta_{\pi}/h$ \cite{Suppl}.

\begin{figure}[ptb]
\includegraphics[width=1\columnwidth]{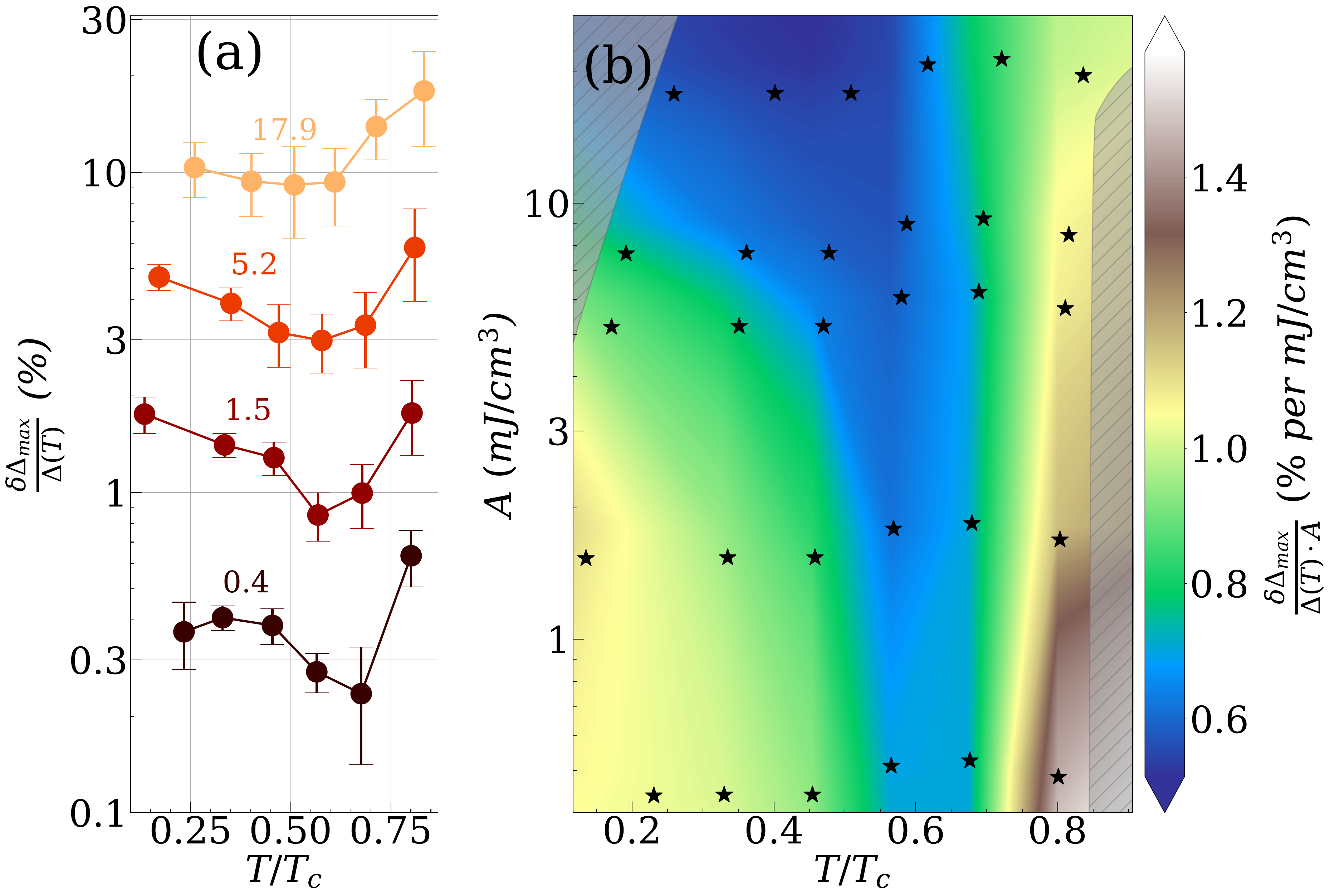}\caption{(a) The amplitude of photoinduced gap suppression for $\nu_{\text{FEL}}=$ 2.7~THz as a function of $T$ for several $A$ (in mJ/cm$^{3}$). (b) All collected data presented in the contour plot versus $A$ and $T$. Star-like marks represent the data points, areas that could not be accessed in the experiment are masked with grey. Note the logarithmic vertical scales. }%
\label{gap_supp}%
\end{figure}

We focus on $T-$ and $A-$dependence of the gap suppression amplitude $\frac{\delta\Delta_{\max}}{\Delta(T)}$ presented in Figure~\ref{gap_supp}. The first observation is a weakly sub-linear dependence of $\frac{\delta\Delta_{\max}}{\Delta(T)}$ on $A$ over the entire $T-$range. This is consistent with the phonon bottleneck model \cite{RothwarfT,RT,DemsarReview,Kondo_Amrit}. Here, following the initial thermalization between condensate(s), QPs and HFPs to a common temperature $T^{\ast}$, part of the absorbed energy is transferred to the HFP subsystem. At low $T$ and $A$ the fraction of energy in the HFP subsystem is negligible \cite{DemsarReview,Kondo_Amrit}. However, this fraction gradually increases with increasing $T^{\ast}$, since $\Delta(T^{\ast})$ reduces and - correspondingly - the specific heat of HFP increases \cite{Kabanov}.

The second, more remarkable observation is that $\frac{\delta\Delta_{\max}}%
{\Delta(T)A}$ decreases with increasing $T$ at constant $A$, with pair-breaking efficiency displaying a pronounced minimum near $T/T_{c}=0.7$. At face value, this result implies that at $T/T_{c}\approx0.7$ the same absorbed energy density results in nearly twice smaller number of broken Cooper-pairs as at the lowest temperatures!

The question arises if the observed decrease in pair-breaking efficiency and its minimum near $0.7$
$T_{c}$ can be accounted for by an increase in the fraction of energy required to heat up HFPs to $T^{\ast}$? To test this hypothesis, we performed simulations
of $\frac{\delta\Delta_{\max}}{\Delta(T)}(T,A)$ using the phenomenological bottleneck model, which is applicable for weak to moderate excitation densities and for temperatures not too close to $T_{c}$ \cite{RT,DemsarReview,Kondo_Amrit}. Assuming a quasi-equilibrium between the QPs and HFPs is reached, their concentrations are determined by the detailed balance equation $Rn_{T^{\ast}}^{2}=\beta N_{T^{\ast}}$, where $n_{T^{\ast}}$ and $N_{T^{\ast}}$ are the QP and HFP concentrations corresponding to a common $T^{\ast}$. The microscopic constants $R$ and $\beta$ are the bare QP recombination rate and the pair-breaking rate by HFP absorption, respectively. Assuming that densities of photoexcited QPs/HFPs are substantially larger than their thermal densities, this boils down to the energy conservation law $A=n_{T^{\ast}}\Delta(T)+2\Delta(T)N_{T^{\ast}}=n_{T^{\ast}}\Delta(T)+2\Delta(T)n_{T^{\ast}}^{2}\frac{R}{\beta}$, with $\Delta(T)$ being the effective gap.

\begin{figure}[h]
\includegraphics[width=1\columnwidth]{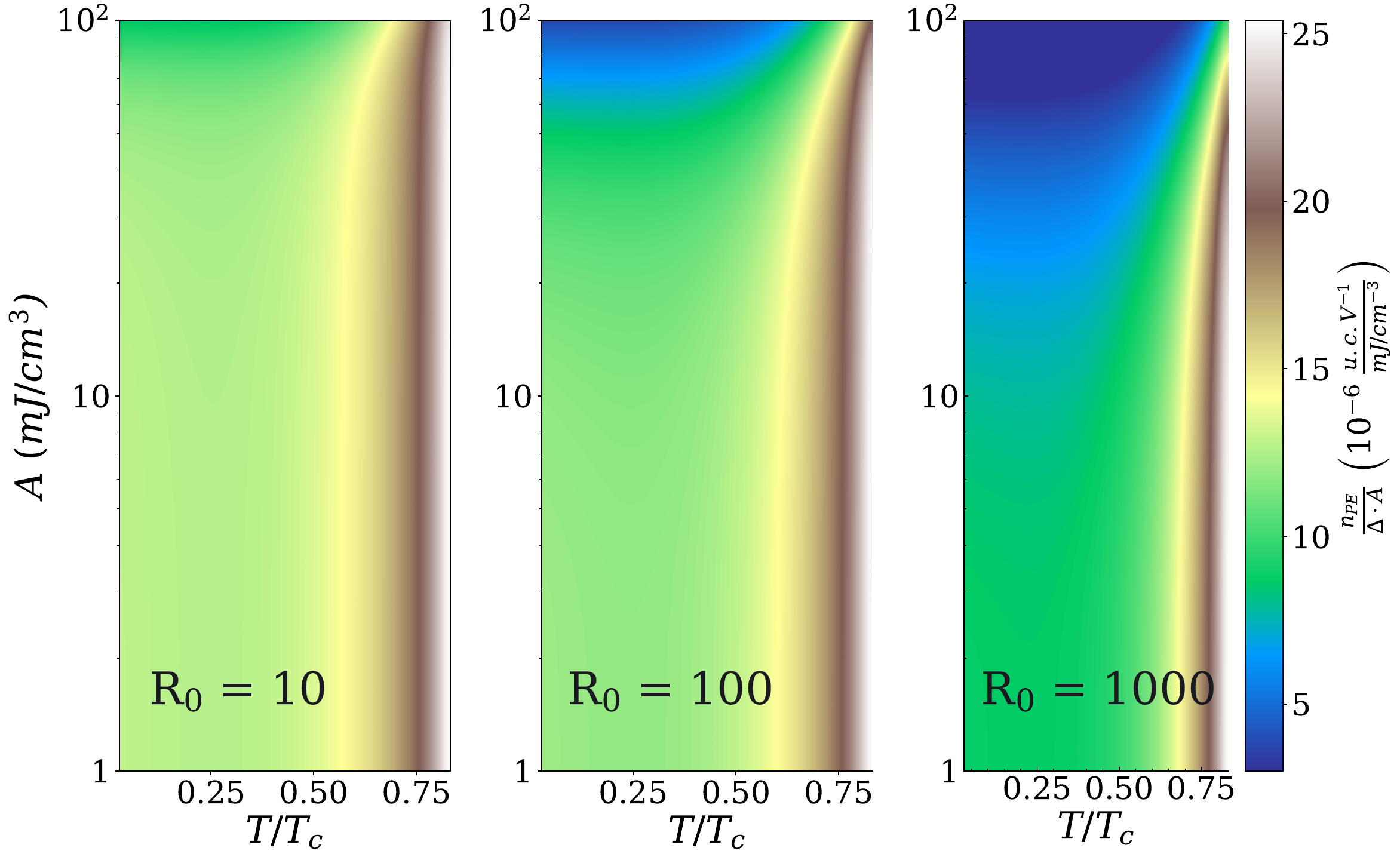}\caption{Simulation of photoinduced gap suppression normalized to $A$ using the phenomenological phonon-bottleneck model. Note the logarithmic vertical scales. The value of QP recombination rate $R$ (in (unit cell volume)/ps) is varied over two orders of magnitude. Simulations reveal the weakly sub-linear $A$-dependence, consistent with experiments. A decrease in pair-breaking efficiency with $T$ (at constant $A$), exhibiting a minimum near $\frac{T}{T_{c}}\approx0.7$ as in see Fig. \ref{gap_supp} is, however, not reproduced.}%
\label{RT_sim}%
\end{figure}

To compare with experiments, we consider that $\delta\Delta$ 
is proportional to the density of photoexcited QPs, $n_{\text{PE}}\approx n_{T^{\ast}}$ and use $\Delta_{\pi}$ as a characteristic gap energy. We further assume $\beta^{-1}\approx30$~ps ($\beta^{-1}\approx15$~ps for NIR excitation \cite{JDMgB2}), and vary $R$ over a large range around $R\approx100$ (unit cell volume)/ps extracted from NIR excitation experiments \cite{JDMgB2}. The results of simulations are presented in Fig.~\ref{RT_sim}. While, similar to experimental results, $\frac{n_{\text{PE}}%
}{A}$ displays a weak sub-linear dependence on $A$ over the entire $T-$range, no decrease or minimum in $n_{\text{PE}}$ with increasing $T$ is observed. Instead, $n_{\text{PE}}$ smoothly increases with $T$ over the entire range of $A$ for all values of $R$.

\begin{figure}[h]
\includegraphics[width=1\columnwidth]{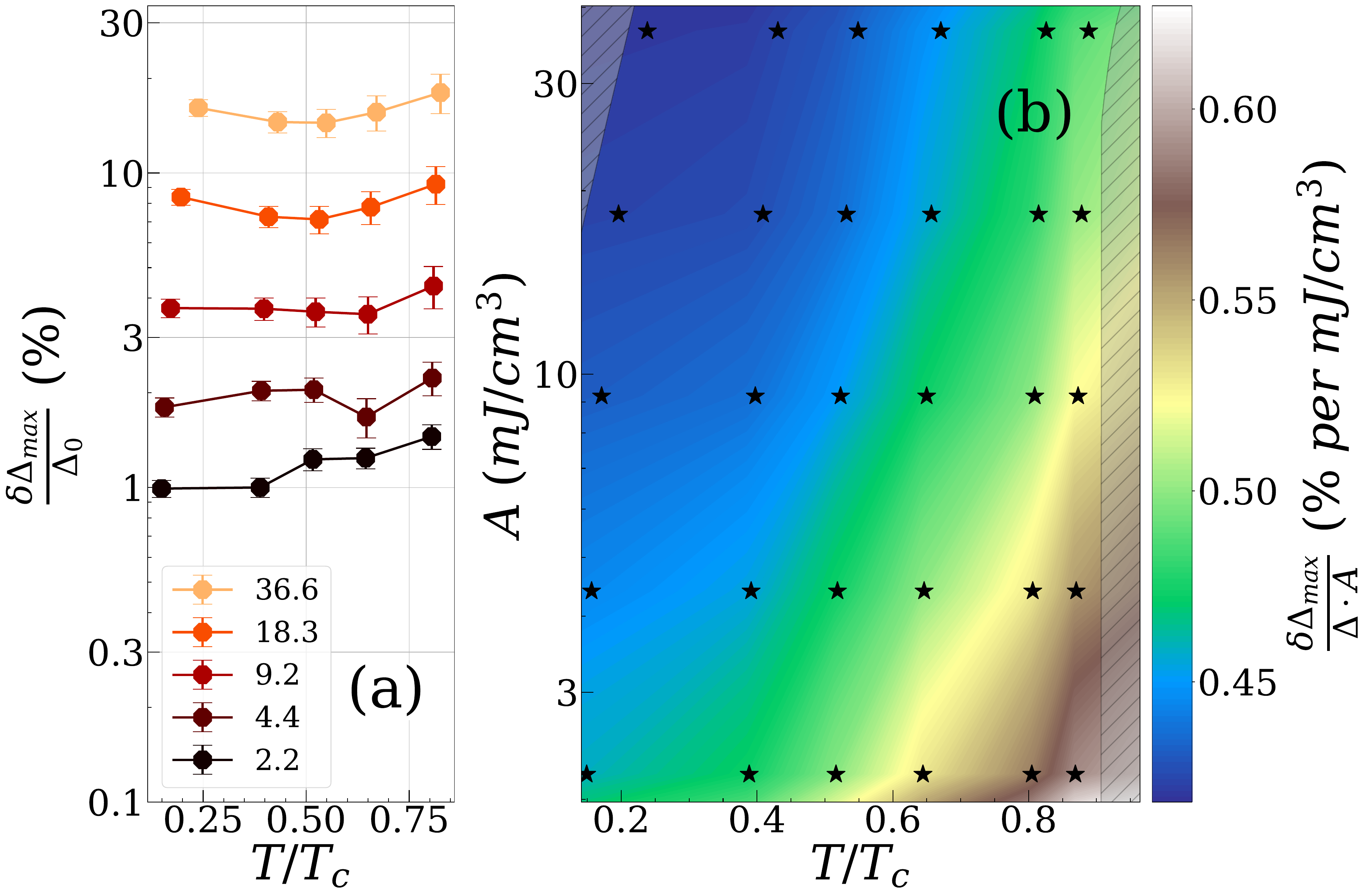}
\caption{(a) The amplitude of gap suppression in case of excitation with NIR pulses as a function of $T$ at different $A$ (in mJ/cm$^{3}$). (b) All data, normalized to $A$, presented in a contour plot. The vertical scales are logarithmic. Star-like markers represent the data points, areas that are not accessible are masked with grey.}%
\label{gap_supp_OPTP}%
\end{figure}

We performed complementary optical pump -- THz probe (OPTP) experiments, where MgB$_{2}$ films were excited with 50~fs NIR pulses at 800~nm. As opposed to THz excitation, NIR photons excite electrons from far below to far above the Fermi energy. In this case, SC is suppressed dominantly via pair-breaking by HFPs created
during the relaxation of hot QPs \cite{DemsarReview}. As the hot carrier relaxation requires a cascade of scattering events, the resulting QP distribution should be much closer to thermal than in the case of narrow-band THz excitation. The data obtained in the OPTP experiment were processed using the same approach as the data obtained with narrow-band THz excitation \cite{Suppl}. The
extracted $\frac{\delta\Delta_{\max}}{\Delta(T)}(A,T)$ for the case of NIR
excitation is shown in Figure~\ref{gap_supp_OPTP}. Unlike in the case of THz pumping (Fig.~\ref{gap_supp}), NIR excitation results in a monotonic increase of $\frac{\delta\Delta_{\max}}{\Delta(T)A}$ with increasing $T$, much like simulated (Fig. \ref{RT_sim}).

The comparison of gap dynamics in MgB$_{2}$ driven by the
THz and NIR pulses demonstrates a substantial reduction in
the pair-breaking efficiency with increasing $T$ under narrow-band THz excitation, with a pronounced minimum near $\approx0.7$ $T_{c}$. The unexpected reduction in the pair-breaking efficiency with $T$ suggests the existence of a process that is competing with THz-driven pair-breaking, becoming stronger as $T$ increases. This observation is consistent with early studies of microwave enhancement of SC, attributed to excitation of thermal QPs away from the gap edge. This scenario also accounts for the observation of a minimum in pair-breaking efficiency near $\approx0.7$ $T_{c}$. Namely, at low-$T$ where $h\nu_{\text{FEL}}<2\Delta_{\sigma}$ no direct pair-breaking in the $\sigma$-band is possible. Near $0.75$ $T_{c}$, $h\nu
_{\text{FEL}}\approx 2\Delta_{\sigma}$ (Fig. \ref{traces_freqs}d), opening an additional pair-breaking channel and thereby increasing the pair-breaking efficiency for $h\nu_{\text{FEL}}\geq2\Delta_{\sigma}$.

Our results suggest Eliashberg-type superconductivity enhancement effects are competing with pair-breaking in THz-driven MgB$_{2}$, implying a highly non-thermal QP distribution on the timescale of at least a few tens of picoseconds. The questions arise, which processes limit the QP thermalization (in the strict sense of QPs reaching the Fermi-Dirac distribution) and what is the difference between optical and THz excitation? As opposed to THz-drive, optical excitation creates QPs with energies in the eV range. On the fs timescale these hot QPs first relax via e-e collisions. When QP energies are reduced down to $\sqrt{\Omega_{D}E_{f}}$, $\Omega_D$ being the Debye energy, e-ph relaxation starts to dominate \cite{VVK}, resulting in a high concentration of phonons on the (sub-)ps timescale \cite{VVK,Novko}. The cascade of scattering events and the multitude of scattering channels, including anharmonic decay of high energy optical phonons\cite{Novko}, is the basis for nearly thermal QP distribution following optical excitation. In the case of THz excitation, thermalization of QPs with energies in the 10 meV range is governed by the e-ph scattering, and can easily take as long as 100 ps at low$-T$ \cite{VVK,Book}. Moreover, due to the coherence factors, the e-ph scattering rate is substantially lower than the QP recombination rate \cite{Kaplan}, the latter not leading to QP thermalization.

In the early studies with continuous microwave irradiation \cite{Clarke1,Clarke2,Tinkham,DemsarReview} the SC enhancement effects of the order of 1\% were observed. Here, using intense THz pulses, the observed reduction in the pair-breaking efficiency is as high as a factor of $\approx2$ at $T\approx0.7$ $T_{c}$. Based on the magnitude of the effect, one may expect that in the absence of pair breaking, \textit{i.e.}, for $h\nu_{\text{FEL}}<2\Delta_{\pi}$, a gap increase of up to a factor of $\approx2$ could be realized at temperatures where the density of thermally excited QPs becomes substantial.

To summarize, the systematic study of SC state dynamics in the two-band BCS superconductor\ MgB$_{2}$\ reveals a pronounced coupling between the $\pi-$ and $\sigma-$band condensates, where the two adiabatically follow each other on a picosecond timescale. At low-$T$, we demonstrate that pair-breaking dynamics depends on $T$, $A$ and $h\nu_{\text{FEL}}$ \cite{Suppl}. While we cannot exclude that the behavior is linked to the coupling between the two condensates, the dependence on $h\nu_{\text{FEL}}$ \cite{Suppl} suggests it is related to the delayed pair-breaking by phonons with $\hbar\omega>2\Delta_{\pi}$ generated when $h\nu_{\text{FEL}}>4\Delta_{\pi}$. Most importantly, tracking the excitation and $T-$dependence of the THz-driven gap reduction, we observe a dramatic decrease in pair-breaking efficiency with increasing $T$, resulting in a pronounced minimum near 0.7 $T_{c}$. The lack of such behavior under photoexcitation with NIR pulses suggests that narrow-band THz excitation results in a highly non-thermal QP distribution on the 100~ps timescale, resulting in Eliashberg-type superconductivity enhancement effects.

\begin{acknowledgments}
This work was funded by the German Research Foundation, Grant No. TRR 288-422213477 (project B08) and TRR 173-268565370 (project A05). T.D. acknowledges support from Alexander von Humboldt Foundation.
\end{acknowledgments}


\begin{thebibliography}{48}
\bibitem [{com()}]{comment}%
  \BibitemOpen
  \href@noop {} {\bibinfo {title} {We consider a generalized version of the
  {T}$^{\ast}$-model, i.e, the {Rothwarf-Taylor} model \cite{RothwarfT},
  applicable also for non-thermal {QP}/high-frequency-phonon distributions.
  {Experimentally}, since the superconductivity-induced changes in the {THz}
  transmission of a {SC} film are governed by the inductive response of the
  condensate \cite{Beck1}, the {THz} transmission presents a direct probe of the
  condensate density.}}\BibitemShut {Stop}%
\end{thebibliography}
\end{document}